\documentclass[preprint]{ptephy_v1}%%%%%% to generate preprint number with ptep logo

\preprintnumber{XXXX-XXXX} %%% %%% Insert preprint number here
\usepackage{hyperref}
\usepackage{pifont}

\usepackage{graphics} %for dealing with figures.
\begin{document}

\title{Gadolinium concentration measurement with an atomic absorption spectrophotometer}

%%%% To generate auto affiliation numbers please use \author{}\affil{} command

\author{Ll. Marti}
\affil{Department of Physics, Yokohama National University, Yokohama, Kanagawa, 240-8501, Japan
\email{martillu@suketto.icrr.u-tokyo.ac.jp} }

\author[2]{~L. Labarga}
\affil{Department of Theoretical Physics, University Autonoma Madrid, 28049 Madrid, Spain}

%\author{Insert third author name here}
%\author[3]{Insert fourth author name here} %%% Use optional bracket [3] to change the respective address
%\affil{Insert third author address here}

%\author{Insert last author name here\thanks{These authors contributed equally to this work}}
%\affil{Insert last author address here}

%%% To include the collaborator name... Please use the command "\collaborator"
%%% For example: \collaborator{ATLAS Collaboration}

\begin{abstract}%
Because gadolinium (Gd) has the highest thermal neutron capture cross section, resulting in an 8 MeV gamma cascade upon capture, it has been proposed for dissolution in water Cherenkov detectors to achieve efficient neutron tagging capabilities. While metallic Gd is insoluble in water, several compounds are very easy to dissolve. Gadolinium sulfate, Gd$_2$(SO$_4$)$_3$, has been thoroughly tested and proposed as the best candidate. Accurate measurement of its concentration, free of doubt from impurities in water, is crucial. An atomic absorption spectrophotometer (AAS) is a device that suits this purpose and is widely used to measure the concentration of many elements. In this study, we describe three different approaches to measure Gd sulfate concentrations in water using an AAS: doping samples with potassium and lanthanum, and employing tantalum and tungsten platforms.
\end{abstract}

\subjectindex{xxxx, xxx}

\maketitle

\section{Introduction and Motivation}
\label{sec:intro}

Water Cherenkov (WC) detectors are very massive and can reconstruct charged particle tracks over a wide energy range. However, their efficiency in detecting neutrons is relatively very low. When neutrons are produced in a WC detector, they first undergo thermalization and are then mostly captured on protons due to the larger thermal neutron capture cross section of protons compared to oxygen nuclei: 0.3 barns and 0.19 millibarns, respectively. Neutron capture on protons results in the production of a single 2.2 MeV gamma within approximately 200 $\mu$s, which is difficult to detect due to its relatively low energy. This single gamma produces few Cherenkov photons to be detected in a WC detector such as Super-Kamiokande~\cite{Zhang:2013tua}.

In 2003, GADZOOKS! was proposed as an approach to achieve efficient neutron tagging in WC detectors~\cite{Beacom:2003nk}. The method involves introducing a gadolinium (Gd) solute into the pure water of WC detectors. The thermal neutron capture cross section of naturally occurring Gd ($\sim$49,000 barns) is about 5 orders of magnitude larger than that of hydrogen. The capture on Gd yields an 8 MeV gamma cascade, which is usually shared among three or four gammas. This technique has been shown to be feasible by diluting Gd sulfate octahydrate, ${Gd}_2(SO_4)_3 \cdot 8H_2O$, as demonstrated at EGADS~\cite{MARTI}. Gd sulfate octahydrate is easy to dissolve, essentially transparent to Cherenkov light, and can be produced to the required radio-purity levels for WC detectors~\cite{hosokawa2022development}. %Additionally, Gd can be easily removed using ion-exchange resins if needed.

\begin{figure}[htb]
\centering
\includegraphics[height=1.9in]{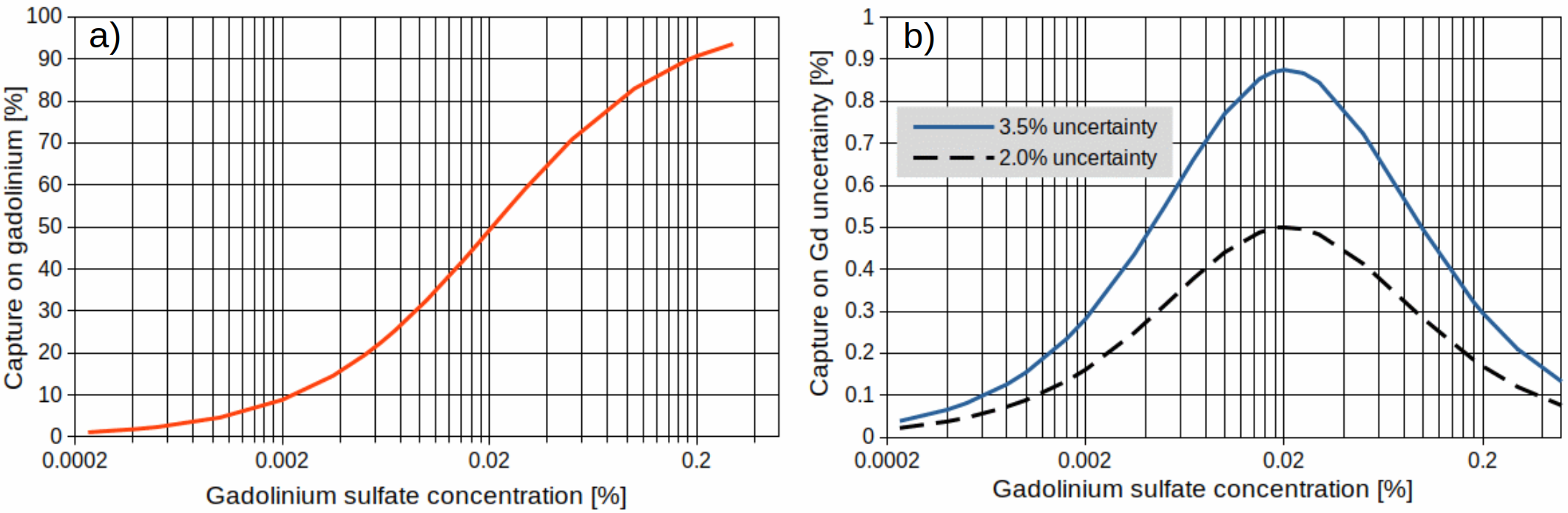}
\caption{Left (a) fraction of neutrons captured on Gd as a function of the Gd concentration in water by mass. Right (b) uncertainty on Gd captures for a 3.5 and 2.0 $\%$ uncertainty in the measurement on the Gd concentration.}
%neutron capture on Gd uncertainty assuming a 3.5 $\%$ uncertainty in the determination of the Gd concentration as a function of the Gd concentration in water by mass.}
\label{fig:captures-and-uncertainty-vs-concentration}
\end{figure}

The concentration of Gd sulfate octahydrate in water, herafter Gd sulfate if not stated otherwise, becomes a critical parameter to measure, as the neutron capture detection efficiency depends on the fraction of neutron captures on Gd. Figure~\ref{fig:captures-and-uncertainty-vs-concentration}a shows this relationship. At a concentration of only 0.02$\%$ (200 ppm) of Gd sulfate, about 50$\%$ of neutron captures are on Gd. At a concentration of 0.2$\%$ (2000 ppm), the fraction of captures on Gd rises to about 90$\%$ while reducing the capture time from 200 $\mu$s to about 30 $\mu$s.

%%%%%%%%%%%%%%%%%%%%%%%%%%%%%%%%%%%%%%%%%%%%%%%%%%%%%%%%%%%%%%%%%%%%%%%%%
%%%%%%%%%%%%%%%%%%%%%%%%%%%%%%%%%%%%%%%%%%%%%%%%%%%%%%%%%%%%%%%%%%%%%%%%%%

Indirect methods such as measuring water resistivity or light absorption in Gd-loaded water can be employed to determine the concentration of Gd sulfate in WC detectors. However, whenever it is necessary to ensure that other elements do not interfere with these measurements, a more reliable method is required, as was the case for the EGADS R$\&$D studies~\cite{MARTI} and later at Super-Kamiokande~\cite{Abe_2022} for its SKGd phases. An atomic absorption spectrophotometer (AAS) is a commonly used device for concentration measurements of a wide range of elements in liquids.

\begin{figure}[htb]
\centering
\includegraphics[height=1.2in]{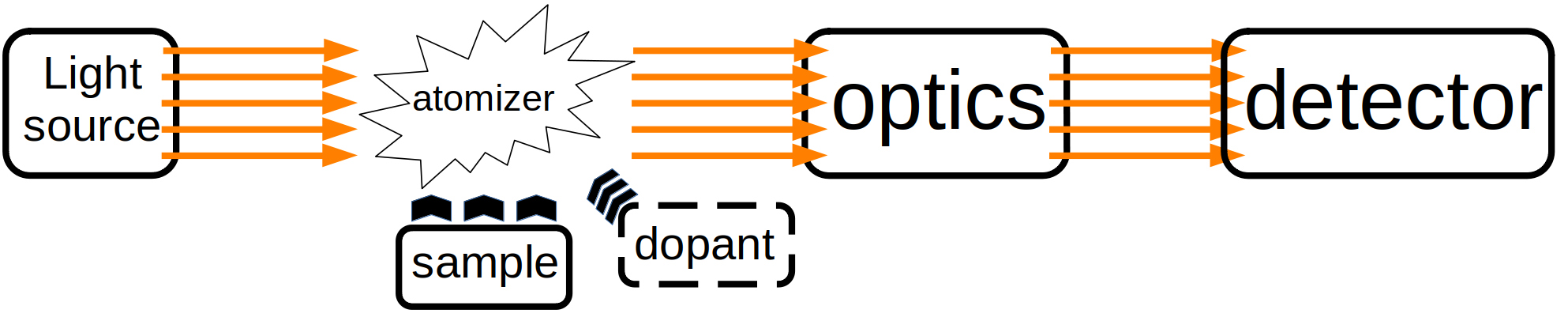}
\caption{Schematic view of an AAS: light source, sample, dopant (optional), atomizer, optics and detector. In our case, the light source is a hollow-cathode lamp made of Gd. To atomize the sample, it is inserted into a cuvette, which may sometimes include a dopant or be a modified cuvette, see text for more details. The detector is a photo-multiplier.}
\label{fig:AAS-schematic-view}
\end{figure}

Figure~\ref{fig:AAS-schematic-view} shows the schematic view of an AAS. A sample of about 15 $\mu$L is inserted into a cuvette, which is typically a container with a cylindrical shape. In some cases a dopant to enhance signal is added as well, see Section~\ref{sec:doping}. Cuvettes are made of pyrolytic carbon and they are sometimes modified  to improve the atomization of the sample by adding a L'vov platform inside them~\cite{LVOV1978153} or variations of this idea, for example in~\cite{liang1991determination}, see Section~\ref{sec:tantalumtungsten}. The sample is then heated to high temperatures to achieve its atomization. During this atomization the sample is illuminated by a hollow-cathode lamp made of the same chemical element of study (in our case Gd) and the absorbance measured with a photo-multiplier as main detector element. The absorbance is a measure of the concentration of the chemical element in the sample. The advantages are evident: by using a Gd hollow-cathode lamp to illuminate an atomized sample, the Gd concentration of a sample can be unequivocally determined. In addition, these measurements require only small sample sizes. %Without sample concentration, i.e., the successive injection and drying of samples before atomization and measurement, the minimum detectable amount of Gd is about 1 ppm.

As shown in Figure~\ref{fig:captures-and-uncertainty-vs-concentration}a the slope of neutron captures on Gd is highest around 0.02$\%$. At this concentration, and considering uncertainties of 3.5$\%$ and 2.0$\%$ in a single Gd sulfate concentration measurement (see Section~\ref{sec:results} for typical values), the uncertainty in neutron captures on Gd are about 0.87$\%$ and 0.50$\%$, respectively. One of our goals is to improve the uncertainty of our measurements.

%%%%%%%%%%%%%%%%%%%%%%%%%%%%%%%%%%%%%%%%%%%%%%%%%%%%%%%%%%%%%%%%%%%%%%%%%
%%%%%%%%%%%%%%%%%%%%%%%%%%%%%%%%%%%%%%%%%%%%%%%%%%%%%%%%%%%%%%%%%%%%%%%%%%

Three main drawbacks were identified for the measurements with an AAS :

\begin{itemize}
\item \textbf{Limited measurement range.} The optimal measurement range for Gd is around 5-30 ppm. However, both the lower and upper bounds of the measurement range are limited by imperfect atomization, as previously described. To measure higher concentrations, precise dilution is necessary, which further increases an already long measurement time and complexity.

\item \textbf{Memory effects.} The AAS atomizes samples at 2700 - 2800 $^\circ$C in a pyrolytic carbon cuvette. While this is sufficient for many elements, similarly to other rare earth metals, Gd requires a higher temperature~\cite{Johansson_1975}. This means that Gd is not fully atomized at these temperatures, reducing the sensitivity of the measurement and leaving some Gd in the cuvette after atomization. Additionally, Gd may react with the carbon of the cuvette and become temporarily fixed, this not only reducing the efficiency of the current atomization but also leading to inaccurate results in subsequent measurements.

Standard AAS measurement methods are susceptible to these atomization inefficiencies - also known as memory effects - which negatively impact the accuracy and precision of Gd concentration measurements using an AAS.

\item \textbf{Cuvette aging.}
The sample is injected into the cuvette, and atomization is achieved through high temperatures. Although cuvettes are designed to withstand the high temperatures needed for atomization, they undergo aging, which can significantly affect the atomization efficiency. The rate of aging is dependent not only on the cuvette type but also exhibits variations among cuvettes of the same type.
\end{itemize}

While AAS is commonly used to measure the concentration of a wide range of elements, these limitations highlight the necessity of a careful reconsideration when using this technique for Gd measurements.
%Although AAS is commonly employed to measure the concentration of a wide range of elements, these limitations  highlight the necessity of careful consideration when using this technique for Gd measurement.

In this paper we present and discuss the different methods developed for the use of an AAS as the technique for the measurement of Gd sulfate concentrations for GADZOOKS! in EGADS and SKGd. Section~\ref{sec:measurement_methods} will describe our measurement methods, including some of the key elements of our apparatus. Section~\ref{sec:results} will present the characterization of the AAS cuvette types and the measurement errors will be derived. The paper will be concluded with a summary of the methods and the main results in Section~\ref{sec:conclusion}.

\section{Measurement Methods}
\label{sec:measurement_methods}

In the following subsection, the general settings common to the different methods will be described, while specific details for each method will be provided in the subsequent subsections.

\subsection{Instrumentation and General Settings}
\label{sec:generalSettings}

For this study, a Hitachi polarized Zeeman AAS of the ZA3000 series was utilized (property of the Institute for Cosmic Ray Research University of Tokyo). Two types of cuvettes were employed: Pyro Tube C HR (used for comparison purposes as its production has been discontinued) and Pyro Tube HR. Hereafter, they will be referred as C HR and HR, respectively. A negative feature shared by both types of cuvettes is that the high temperatures required for sample atomization or cuvette cleaning gradually degrade their quality, although, as it will be shown, to varying degrees. Henceforth, these two processes will be referred to as furnace processes. The differences seen in the cuvette types will emphasize the importance of cuvette type availability and choice. There is a limited number of times these processes can be repeated before their effectiveness significantly diminishes, thereby determining their effective lifetime. Based on our experience, C HR cuvettes could be used for almost 100 furnace processes, whereas HR cuvettes can be used for only about 70. 

Gd hollow-cathode lamps were operated with a current of 10 mA. The analysis wavelength was set to 422.6 nm, and the slit width was adjusted to 1.3 nm to maximize sensitivity for the AAS photo-multiplier and other relevant conditions. Concentrations were determined based on the measured absorption\footnote{Strictly speaking, it measures absorbance.} peak height (ABS). Argon (Ar) gas was supplied to aid in sample drying and to protect the sample from being contaminated and the cuvette from aging at high temperatures. After turning on the Gd hollow-cathode lamp, about 5 minutes are needed to warm it up and ensure light intensity stability well within 1$\%$.

%%%%%%%%%%%%%
\subsection{Measurement General Settings}
\label{sec:measmeth}

This paper focuses on two types of measurements: doping with lanthanum (La) and potassium (K), and using platforms made of either tantalum (Ta) or tungsten (W). These platforms are inserted inside the cuvette before measurement start. As it will be shown, doping with La and K (La+K) requires more furnace processes compared to using Ta/W platforms. Consequently,  doping becomes increasingly impractical with HR cuvettes due to the limited number of furnace processes this cuvette type can perform. As a result, when the production of C HR cuvettes was discontinued, we were compelled to search for a better solution. The measurement methods and cuvettes used are summarized in Table~\ref{tab:cuvette-vs-method}.

%%%%%%%%%%%%%%%%%%%%%%%%%%%%%%%%%%%%%%%%%%%%%%%%%%%%%%%%%%%%%%%%%%%%%%%%%
\begin{table}[h]
\vspace{-0.2cm}
\begin{center}
\renewcommand\arraystretch{1.1}  % vertical   separation
\renewcommand{\tabcolsep}{5pt}   % horizontal separation
%\newcolumntype{C}{>{\centering\arraybackslash}m{1cm} }
\begin{tabular}{|c|c|c|c|c|c|c}  
\hline
             & Doping with  &  Ta and W       \\
             & La+K         &  platforms      \\ \hline
 Pyro Tube C HR  & \checkmark   &  \ding{55}             \\ \hline
 Pyro Tube HR    & \checkmark   &  \checkmark            \\
\hline
\end{tabular}
\caption{Table with the different methods and cuvette types employed.}
\label{tab:cuvette-vs-method}
\end{center}
\end{table}
%%%%%%%%%%%%%%%%%%%%%%%%%%%%%%%%%%%%%%%%%%%%%%%%%%%%%%%%%%%%%%%%%%%%%%%%%%%

Once the AAS has been prepared, see Section~\ref{sec:generalSettings}, the basic sequence begins. Their common settings (dry stages and ash) are shown in Table~\ref{tab:temp_settings-common}. %The three first stages: dry-1, dry-2 and ash are common to all methods and cuvettes. 
The settings for the subsequent stages (atomize, heating and cool) differ depending on the cuvettes and method. Specifically, the settings for C HR cuvettes and doping with La+K are shown in Table~\ref{tab:temp_settingsC_HR}, while the settings for HR cuvettes are shown in Tables~\ref{tab:temp_settings-KLaHR} and~\ref{tab:temp_settings-platformHR}, corresponding to the La+K doping and Ta/W platform methods, respectively.

%%%%%%%%%%%%%%%%%%%%%%%%%%%%%%%%%%%%%%%%%%%%%%%%%%%%%%%%%%%%%%%%%%%%%%%%%
\begin{table}[h]
%~~~~~~~~~Temperature units in $^\circ$C, time intervals in seconds and flow in mL/min.
\vspace{-0.2cm}
\begin{center}
\renewcommand\arraystretch{1.1}  % vertical   separation
\renewcommand{\tabcolsep}{5pt}   % horizontal separation
%\newcolumntype{C}{>{\centering\arraybackslash}m{1cm} }
\begin{tabular}{|c|c|c|c|c|c|c}  
\hline
 Program    & Start        &  End          & Ramp  & Hold   &  Ar   \\
 stage      & temperature  &  temperature  & time  & time   & flow  \\ \hline
 Dry-1      & 50           &  130          & 70    &  -     & 200   \\ \hline
 Dry-2      & 130          &  500          & 30    &  -     & 200   \\ \hline
 Ash        & 1800         &  1800         & 20    &  -     & 200   \\ \hline
\end{tabular}
\caption{The three first stages: dry-1, dry-2 and ash of the basic sequence. They are common to all methods and cuvettes. See text for details. Temperature units in $^\circ$C, time intervals in seconds and flow in mL/min.}
\label{tab:temp_settings-common}
\end{center}
\end{table}
%%%%%%%%%%%%%%%%%%%%%%%%%%%%%%%%%%%%%%%%%%%%%%%%%%%%%%%%%%%%%%%%%%%%%%%%%%%

%%%%%%%%%%%%%%%%%%%%%%%%%%%%%%%%%%%%%%%%%%%%%%%%%%%%%%%%%%%%%%%%%%%%%%%%%
\begin{table}[h]
%~~~~~~~~~Temperature units in $^\circ$C, time intervals in seconds and flow in mL/min.
\vspace{-0.2cm}
\begin{center}
\renewcommand\arraystretch{1.1}  % vertical   separation
\renewcommand{\tabcolsep}{5pt}   % horizontal separation
%\newcolumntype{C}{>{\centering\arraybackslash}m{1cm} }
\begin{tabular}{|c|c|c|c|c|c|c}  
\hline
 Program    & Start        &  End          & Ramp  & Hold   &  Ar   \\
 stage      & temperature  &  temperature  & time  & time   & flow  \\ \hline
 Atomize    & 2800         &  2800         & -     & 10     &  10   \\ \hline
 Heating    & 3000         &  3000         & -     & 10     & 200   \\ \hline
 Cool       &  -           &   -           & -     & 20     & 200   \\ \hline
\end{tabular}
\caption{Settings for measurements with C HR cuvettes and doping with La+K. The photo-multiplier voltage was set to 260 V. Sample injection volume: 15 $\mu$L. After the cooling stage, 25 $\mu$L of pure water are injected in the cuvette and the stages of heating and cooling are repeated twice. See text for details. Temperature units in $^\circ$C, time intervals in seconds and flow in mL/min.}
\label{tab:temp_settingsC_HR}
\end{center}
\end{table}
%%%%%%%%%%%%%%%%%%%%%%%%%%%%%%%%%%%%%%%%%%%%%%%%%%%%%%%%%%%%%%%%%%%%%%%%%%%

%%%%%%%%%%%%%%%%%%%%%%%%%%%%%%%%%%%%%%%%%%%%%%%%%%%%%%%%%%%%%%%%%%%%%%%%%
\begin{table}[h]
%~~~~~~~~~Temperature units in $^\circ$C, time intervals in seconds and flow in mL/min.
\vspace{-0.2cm}
\begin{center}
\renewcommand\arraystretch{1.1}  % vertical   separation
\renewcommand{\tabcolsep}{5pt}   % horizontal separation
%\newcolumntype{C}{>{\centering\arraybackslash}m{1cm} }
\begin{tabular}{|c|c|c|c|c|c|c}  
\hline
 Program    & Start        &  End          & Ramp  & Hold   &  Ar   \\
 stage      & temperature  &  temperature  & time  & time   & flow  \\ \hline
 Atomize    & 2800         &  2800         & -     & 10     &  10   \\ \hline
 Heating    & 2800         &  2800         & -     & 10     & 200   \\ \hline
 Cool       &  -           &   -           & -     & 20     & 200   \\ \hline
\end{tabular}
\caption{Settings for measurements with HR cuvettes and doping with La+K. The photo-multiplier voltage was set to 400 V. Sample injection volume: 15 $\mu$L. After the cooling stage, 25 $\mu$L of pure water are injected in the cuvette and the stages of heating and cooling are repeated twice. See text for details. Temperature units in $^\circ$C, time intervals in seconds and flow in mL/min.}
\label{tab:temp_settings-KLaHR}
\end{center}
\end{table}
%%%%%%%%%%%%%%%%%%%%%%%%%%%%%%%%%%%%%%%%%%%%%%%%%%%%%%%%%%%%%%%%%%%%%%%%%%%

%%%%%%%%%%%%%%%%%%%%%%%%%%%%%%%%%%%%%%%%%%%%%%%%%%%%%%%%%%%%%%%%%%%%%%%%%
\begin{table}[h]
%~~~~~~~~~Temperature units in $^\circ$C, time intervals in seconds and flow in mL/min.
\vspace{-0.2cm}
\begin{center}
\renewcommand\arraystretch{1.1}  % vertical   separation
\renewcommand{\tabcolsep}{5pt}   % horizontal separation
%\newcolumntype{C}{>{\centering\arraybackslash}m{1cm} }
\begin{tabular}{|c|c|c|c|c|c|c}  
\hline
 Program    & Start        &  End          & Ramp  & Hold   &  Ar   \\
 stage      & temperature  &  temperature  & time  & time   & flow  \\ \hline
 Atomize    & 2700         &  2700         & -     & 10     &  10   \\ \hline
 Heating    & 2800         &  2800         & -     & 10     & 200   \\ \hline
 Cool       &  -           &   -           & -     & 20     & 200   \\ \hline
\end{tabular}
\caption{Settings for measurements with HR cuvettes and Ta/W platforms. The photo-multiplier voltage was set to 400 V. Sample injection volume: 13 $\mu$L. After the cooling stage, a heating stage of 5 seconds followed by a cooling stage of 20 seconds is enough to ensure no memory effects. See text for details. Temperature units in $^\circ$C, time intervals in seconds and flow in mL/min.}
\label{tab:temp_settings-platformHR}
\end{center}
\end{table}
%%%%%%%%%%%%%%%%%%%%%%%%%%%%%%%%%%%%%%%%%%%%%%%%%%%%%%%%%%%%%%%%%%%%%%%%%%%

The total duration of the basic sequence is about 160 seconds (excluding sample injection time). The ramp time of both drying processes were set to be long enough to ensure a slow and smooth drying process, preventing sample bubbling or spillage within the cuvette. Suppressing bubbling and spillage ensures an efficient and homogeneous atomization resulting in enhanced signal quality and improved repeatability. Despite the absence of organic materials in our samples, the ashing step was included to broaden the applicability of our results and make them relevant for other purposes. Note the higher temperatures the C HR cuvettes can achieve as compared to those with HR cuvettes\footnote{Cuvette maximum temperatures are provided by the manufacturer, Hitachi, Ltd.}. This allows the former to achieve a better atomization of the sample. 

The heating stage following atomization serves as a cleaning process to eliminate any residual impurities in the cuvette. However, this is not achieved when doping with La+K and a cleaning process is repeated twice, independently of cuvette type. This cleaning process consists in the injection of 25 $\mu$L of pure water into the cuvette followed by the heating and cooling stages, similar to those shown in Tables~\ref{tab:temp_settingsC_HR} and~\ref{tab:temp_settings-KLaHR}. Sometimes this procedure may not be sufficient, and a blank measurement, i.e. including all stages, is needed to completely remove all impurities. For the measurements with platforms, a heating stage of 5 seconds followed by a cooling stage of 20 seconds is enough to ensure no memory effects. The difference in the cleaning procedures needed after atomization translates to a much longer measurement time for the doping with La+K method compared to the Ta and W platforms method. Including samples and doping injection time, the measurement time for the Ta and W platforms method is only 25$\%$ of that required by the La+K doping method.

%\subsection{Simple cleaning measurement method}
%\label{sec:simple}
%
%For the analysis of most elements other than Gd, after the cooling phase a new sample concentration measurement can be done. However, when measuring Gd samples, the large memory effects need to be taken cared of first. One simple method to reduce memory effects is to use a cleaning procedure. It consists of inserting a blank sample (pure water) of 25 $\mu$L and going through Clean phase in Table~\ref{tab:temp_settings}. With Pyro Tube C HR cuvettes, this had to be done twice to ensure the reduction of the memory effects, so they could be neglected. With the Pyro Tube HR cuvettes, however, the cleaning procedure had to be done three times before proceeding with the next sample measurement. This further increased the measurement time and added another atomization needed for each sample measurement, an additional reducion its already limited number of atomizations. 

\subsection{Doping with Potassium and Lanthanum}
\label{sec:doping}

%Akira Yonetani, Hitachi

To increase the signal and reduce memory effects two dopants were found to be useful~\cite{personal_communication}: lanthanum (La)~\cite{AN9608500495, YOFE1958166} and potassium (K)~\cite{SANUI1968330, WILLIS1960259}.

AAS measurements rely on the absorption of light from the hollow-cathode lamp by the ground state atoms of an element, in our case Gd. Ionized Gd atoms do not absorb radiation like neutral ground states do. It has been shown that, under similar conditions, adding K, which has a low ionization potential, increases the signal, i.e. the ABS of light by the measured element~\cite{SANUI1968330}.

When analysing sulfates it has been found the presence of chemical interference that can distort the measurement results. It has been reported that La can suppress this interference under similar conditions~\cite{AN9608500495, YOFE1958166}. The formation of carbides, compounds of carbon and a metal, hinders atomization~\cite{liang1991determination}, making it an important effect to avoid. Applying a coating of lanthanum (La) is known to reduce this effect~\cite{ebdon1991introduction}.

%The usage of La and K as dopants to measure Gd sulfate will be reported here. 
Approximately 15 $\mu L$ of the sample are injected into the cuvette for the measurement using this method. The La+K doping is injected separately. The best results were obtained when injecting 10 $\mu L$ of a mixture containing K and La in a proportion of 100:5 from 1000 ppm solutions.

\subsection{Tantalum and Tungsten Platforms}
\label{sec:tantalumtungsten}

% size of Ta and W chips w = 4.37 mm, l = 7.50 mm
% diameter of mould pipe: 4.35 mm
% thickness of Ta and W foils 0.2, 0.25 and 0.3 mm 

Because the production of C HR cuvettes was discontinued and the serious limitations of the HR cuvettes (lower number of furnace processes during its lifetime and maximum temperature), a new method was required to reduce memory effects. We drew inspiration from L'vov platforms and a previous study that utilized Ta foils~\cite{liang1991determination}, although we arrived at somewhat different results. Ta is a transition metal that exhibits excellent heat conductivity, low thermal expansion, and high melting point (3017$^\circ$C) and density (16.7 g/cm$^3$). Ta is a rather ductile metal that is relatively easy to work with. We searched for similar or even better candidates than Ta and found that W was a suitable candidate due to its higher melting point (3422$^\circ$C) and density (19.25 g/cm$^3$), which are comparable to those of uranium and gold. Considering that these platforms are inserted into the cuvette, the low expansion coefficient of W, which is the lowest among all pure metals, contributes to achieving mechanical stability inside the cuvette.

These platforms were made by machining foils of Ta and W of 20 cm x 20 cm. Foil thicknesses of 0.1 mm, 0.2 mm and 0.3 mm were evaluated. While platforms of 0.1 mm did not work reliably for more than a few measurements, platforms of 0.2 mm would last for at least one cuvette lifetime. However, platforms with a thickness of 0.3 mm were more reliable and could typically last through the lifetimes of at least two cuvettes\footnote{If the platform showed a crack or any other damage, it would not be used again.}

Various platform sizes were evaluated as well. The best results were obtained with pieces measuring 4.37 mm in width and 7.50 mm in length. Usual machining tools tend to damage the edges of the platforms. Therefore, a company was commissioned to machine these pieces using electrical discharge machining (EDM). Although other techniques such as water jet machining may also be viable. EDM has proved to be an effective, precise and quite economical way of machining Ta and W pieces out of larger foils.

\begin{figure}[htb]
\centering
\includegraphics[height=2.1in]{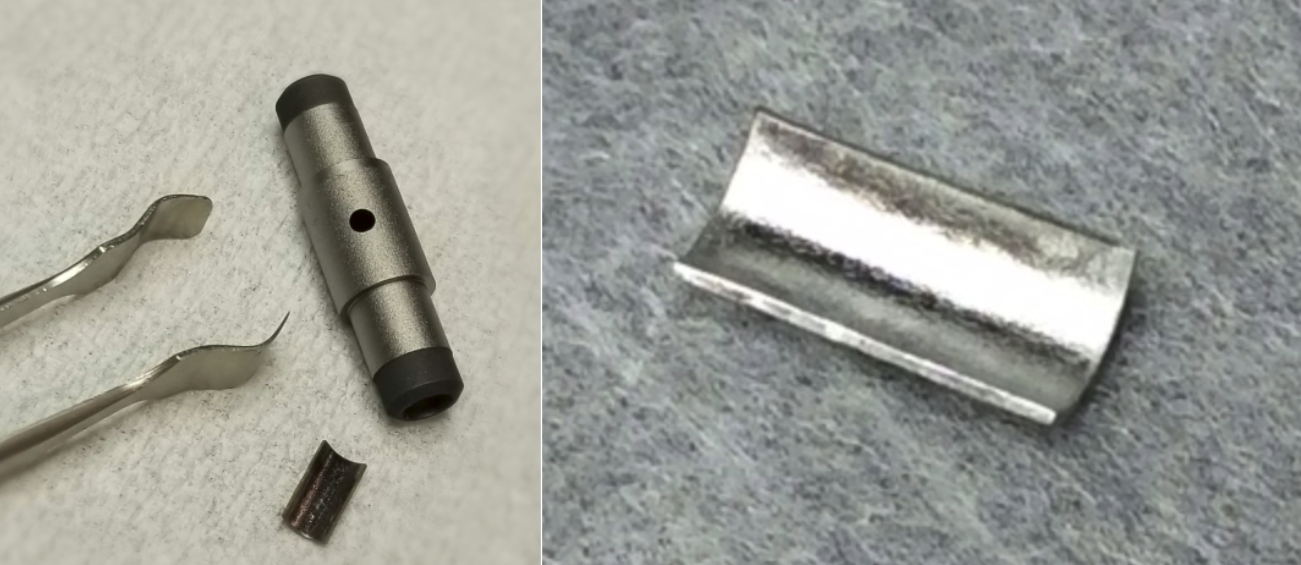}
\caption{Left: HR cuvette and tantalum platform after formation, ready to be inserted into the cuvette. Right: Close-up view of the platform.}
\label{fig:platform_picture}
\end{figure}

Several shapes, from just flat to curved with different curvatures, were tested. The best results were achieved when the pieces were molded into a pipe with a diameter of 4.35 mm, see Figure~\ref{fig:platform_picture}. However, while Ta was easy to work with in the molding process, the brittleness of W posed more challenges. To mold the W pieces they had to be heated to approximately 450$^\circ$C~\cite{KRSJAK201481}. A Hakko FR-810B hot air rework station was utilized to heat the W piece within the mold itself before giving it the final shape.

%%%%%%%%%%%%%%%%%%%%%%%%%%%%%%%%%%%%%%%%%%%%%%%%%%%%%%%%%%%%%%%%%%%%%%%%%
%%%%%%%%%%%%%%%%%%%%%%%%%%%%%%%%%%%%%%%%%%%%%%%%%%%%%%%%%%%%%%%%%%%%%%%%%%

% see presentation on 1/June/2016 (Ta and W properties, Gd meeting)

% CM 18/June/2016 (new methods)
% see presentation on 17/Jan/2013 (new AAS, WWLowE meeting)
% see presentation on 11/May/2013 (uncertainties, collaboration meeting)
% see presentation on 1/Sep/2010 (uncertainties, WWLOWE meeting)
% see presentation on 12/Nov/2010 (uncertainties, "marti10_autumn-3.pdf" collaboration meeting)

The sample injection volume was reduced from 15 $\mu L$ to 13 $\mu L$ to prevent it from spilling outside the Ta/W platforms onto the cuvette floor, which would lead to repeatability issues.

\section{Results}
\label{sec:results}

As stated in the Introduction, Section~\ref{sec:intro}, the Gd concentration is a critical parameter to measure, and our main goal is to improve the uncertainty of our measurements. Before we assess the uncertainties of our methods, we investigate how the average ABS of a sample changes as a function of the number of furnace processes a cuvette has undergone. Additionally, we study the differences from one cuvette to another. This will lead to the determination of a correction factor that is estimated for each cuvette and method, see~\ref{sec:ABSslope}. Using this correction, we then estimate the uncertainties for each cuvette type and method in~\ref{sec:uncertainties}.

\subsection{ABS dependence on number of furnace processes}
\label{sec:ABSslope}

As shown in Tables~\ref{tab:temp_settingsC_HR},~\ref{tab:temp_settings-KLaHR} and~\ref{tab:temp_settings-platformHR}, the cuvettes reach temperatures up to about 2800-3000$^\circ$C. Such temperatures are necessary to achieve an efficient sample atomization and cleaning to remove any impurity left in the cuvette. Prolonged exposure to high temperatures gradually degrades the cuvettes, requiring their replacement at a certain point: approximately 100 atomizations for C HR and about 70 for HR cuvettes.

Measuring the ABS of a sample requires to go through the basic sequence that was detailed in Section~\ref{sec:measmeth}. The ABS of a sample varies after a certain number of furnace processes. This variation depends not only on the total number of furnace processes that the current cuvette has undergone but also on the efficiency of the preceding furnace processes as well. While cuvette aging leads to a decrease in ABS values, atomization inefficiencies or memory effects potentially have the opposite effect, artificially increasing ABS values. Consequently, in the ideal scenario of perfect sample atomization, we would observe solely the degradation of the cuvette, indicated by decreasing ABS values after several furnace processes.

The ABS variation of a sample typically follows a linear trend for all cuvette types and analysis methods. The slope of the ABS variation depends on the cuvette aging rate and the accumulated memory effects. Measuring the slope allows for the correction of these effects and enhances the precision and accuracy of our measurements. Although the rate of quality degradation and the extent of memory effects vary between individual cuvettes, it is interesting to asses the quality traits of the two cuvette types and assess differences among measurement methods in this context. In this context, the significance of both cuvette choice and its availability from the maker, as well as the selected method, will become apparent.

\begin{figure}[htb]
\centering
\includegraphics[height=4.1in]{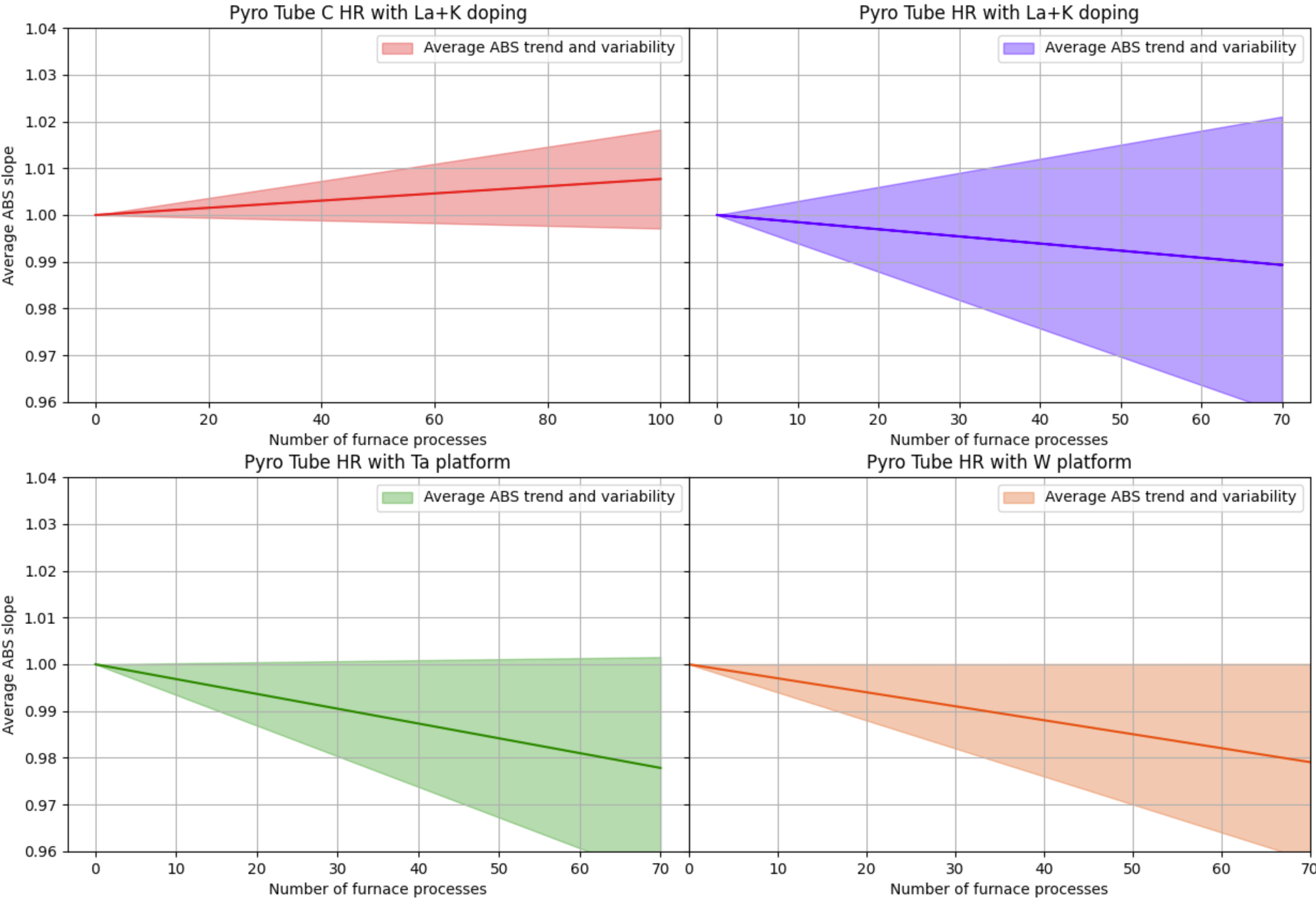}
\caption{Quality traits of cuvettes and analizing methods: average ABS slope of a sample as a function of furnace processes for Pyro Tube C HR cuvettes with La+K doping method (upper left), Pyro Tube HR cuvette with La+K doping method (upper right), Pyro Tube HR cuvette with the Ta platform method (bottom left) and Pyro Tube HR cuvette with W platform method (bottom right). The colored areas show the observed slope variability for a given cuvette type and method.}
\label{fig:cuvette-traits}
\end{figure}

Figure~\ref{fig:cuvette-traits} represents straight lines whose slopes are obtained from fits to sets of about 8 ABS measurements of 20 ppm Gd sulfate standard samples along the cuvette lifetimes (max. 100 for C HR and 70 for HR). This was done for the C HR and HR cuvette types for the methods of La+K doping and using Ta and W platforms. The central line corresponds to the average slope and the colored area spans the observed slope variability.

The C HR cuvettes with the La+K doping method (upper left) exhibit an average positive slope. This means that when comparing from one cuvette to another, the difference in memory effects are the most important factor in the behavior observed for this cuvette type and method combination. The memory effects could have been reduced by adopting a more aggressive cleaning method. We could have tried this if the C HR cuvettes production would not have been discontinued. However, the associated costs in measurement time and increase (decrease) of furnace processes (cuvette lifetime) would have made this undesirable.

An average negative slope is observed for the HR cuvette type with the method of La+K doping (upper right), although the slope variability is large and in many instances, it is positive. This indicates that cuvette aging is more important for this cuvette type than for the C HR type. The slope variability among individual cuvettes is also larger compared to C HR cuvettes.

The HR cuvette type with the Ta platforms method (bottom left) also exhibits an average negative slope, which is more pronounced compared to the previous case. As the cuvette types are the same in these two cases, this suggests that the memory effects have been significantly diminished. The slope variability for this method is lower compared to the case of La+K doping with the same cuvette type. However, we still observe a larger slope variability for this cuvette type and method compared to the C HR cuvette type with La+K doping.

The HR cuvette type with the W platforms method (bottom right) exhibits a similar average slope and variability to that of Ta platforms. Because both the Ta and W platforms typically outlast the lifetime of a single HR cuvette, and their performance is restored when inserted into a second cuvette, this suggests that most of the observed aging can be attributed to the cuvette itself.

%%%%%%%%%%%%%%%%%%%%%%%%%%%%%%%%%%%%%%%%%%%%%%%%%%%%%%%%%%%%%%%%%%%%%%%%%
%%%%%%%%%%%%%%%%%%%%%%%%%%%%%%%%%%%%%%%%%%%%%%%%%%%%%%%%%%%%%%%%%%%%%%%%%%

%->RESUMIR RESULTADO FIGURA
In summary, Figure~\ref{fig:cuvette-traits} shows how in general the ABS of one sample would change during the lifetime of a cuvette for a given method and cuvette type. C HR cuvettes last longer, the ABS of a sample does not change much during a cuvette's lifetime and a slight memory effect is seen. For the HR cuvette type, we observe greater variations from one cuvette to another. With Ta/W platforms the memory effects basically disappear and we are left with the cuvette aging effects. This makes us wonder how the combination of a C HR cuvette and a Ta/W platform would compare if cuvette C HR or an equivalent were available. Note that this figure does not reflect the measured ABS variations of one sample within the lifetime of a cuvette, which we will study next.

\subsection{Measurement Uncertainty}
\label{sec:uncertainties}

\begin{figure}[htb]
\centering
\includegraphics[height=1.85in]{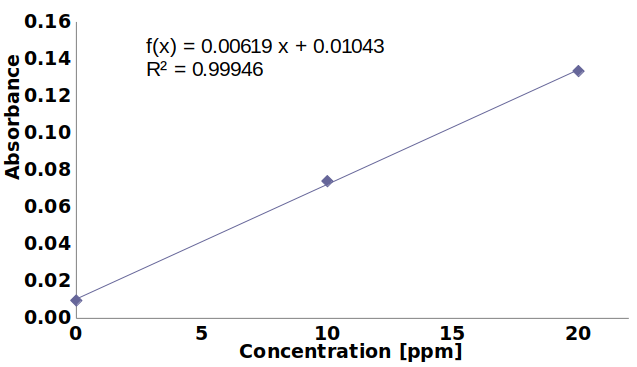}
\caption{Example of calibration data for an HR cuvette with a Ta platform with three Gd sulfate concentrations: pure water (0 ppm), 10 ppm and 20 ppm (linear fit included).}
\label{fig:workingcurve}
\end{figure}

After applying a linear correction for aging and memory effects to a series of samples with known concentration for each cuvette type and measurement method combination, the dispersion of concentration measurements was studied. Before concentration measurements of Gd sulfate samples can be done, a calibration of the AAS is needed to translate from the sample ABS to Gd sulfate concentration. For that, the ABS of 3 standard samples are measured: blank (pure water), 10 ppm Gd sulfate and 20 ppm Gd sulfate. For each sample the ABS is measured twice and the average is used in a linear fit ABS vs concentration. An example of such calibration is shown in Figure~\ref{fig:workingcurve}, specifically for an HR cuvette with a Ta platform. Similar examples, particularly regarding linearity, are obtained for the other cuvettes and methods. After that, from the measured ABS of samples their concentration can be inferred.

%%%%%%%%%%%%%%%%%%%%%%%%%%%%%%%%%%%%%%%%%%%%%%%%%%%%%%%%%%%%%%%%%%%%%%%%%
%%%%%%%%%%%%%%%%%%%%%%%%%%%%%%%%%%%%%%%%%%%%%%%%%%%%%%%%%%%%%%%%%%%%%%%%%%

\begin{figure}[htb]
\centering
\includegraphics[height=3.2in]{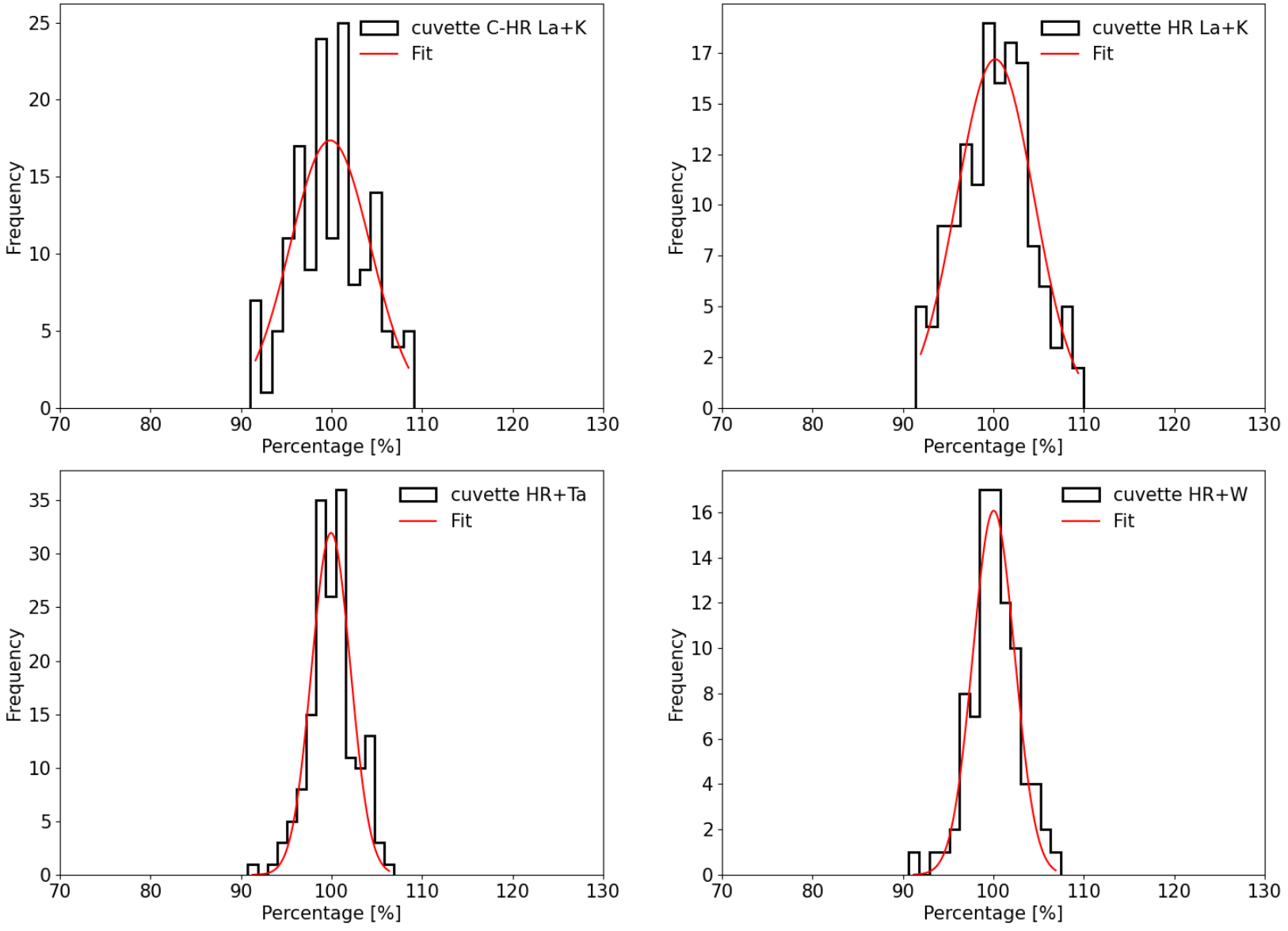}
\caption{Distribution of measured concentration as percentage from the original concentration of standard samples with the Pyro Tube C HR cuvettes with La+K doping method (upper left), Pyro Tube HR cuvette with La+K doping method (upper right), Pyro Tube HR cuvette with the Ta platform method (bottom left) and Pyro Tube HR cuvette with W platform  method (bottom right). These distributions include the correction for the aging and memory effects.}
\label{fig:cuvette-methods-dispersion}
\end{figure}

Figure~\ref{fig:cuvette-methods-dispersion} shows the distributions of singly measured concentrations of samples with known concentration as percentage from their standard value. The concentrations were 10 and 20 ppm Gd sulfate. As preliminary studies by us showed  no significant dependence of the resolution on the concentration measurement, we have used the combined sample. These distributions represent the resolution of our different approaches: C HR cuvettes with La+K doping method, HR cuvettes with La+K doping method, HR cuvettes with the Ta and W platforms method, from top to bottom and from left to right. For the C HR cuvettes with La+K doping method (upper left), a standard deviation of about 5.0$\%$ is observed. This is very similar, also about 4.9$\%$, to the HR cuvettes with La+K doping method (upper right). The dispersion is significantly reduced when using the Ta, with a standard deviation of 2.6$\%$, and W platform, with a standard deviation of 2.8$\%$, methods (bottom left and right, respectively).

%%%%%%%%%%%%%%%%%%%%%%%%%%%%%%%%%%%%%%%%%%%%%%%%%%%%%%%%%%%%%%%%%%%%%%%%%

Note that after correcting for aging and memory effects, the variance of the distribution of measured concentration is significantly smaller for Ta/W platforms than for La+K doping, the later being mostly independent of cuvette type. These results are consistent with a very significant reduction of memory effects when using the Ta/W platforms method even though smaller variations were seen from one C HR cuvette to another.

For our Gd sulfate measurements at EGADS and Super-Kamiokande, two consecutive ABS measurements are performed and combined into an average. This is what we commonly call a sample measurement. Based on the evaluated standard errors mentioned above, the errors for each cuvette type and method can be calculated. These errors are summarized in Table~\ref{tab:cuvette-vs-method-errors}. Since the primary contribution to the errors is of a statistical nature, further reduction in errors can be achieved through subsequent measurements.

%%%%%%%%%%%%%%%%%%%%%%%%%%%%%%%%%%%%%%%%%%%%%%%%%%%%%%%%%%%%%%%%%%%%%%%%%
\begin{table}[h]
    \vspace{-0.2cm}
    \begin{center}
        \renewcommand\arraystretch{1.1}  % vertical separation
        \renewcommand{\tabcolsep}{5pt}   % horizontal separation
        %\newcolumntype{C}{>{\centering\arraybackslash}m{1cm} }
        \begin{tabular}{|c|c|c|c|c|c|c}  
            \hline
            Measurement  & Doping with  & \multicolumn{2}{c|}{Platforms} \\
            \cline{3-4}
            resolution      & La+K      &  \multicolumn{1}{c|}{Ta} & \multicolumn{1}{c|}{W} \\ \hline
            Pyro Tube C HR  & 3.5$\%$   & \multicolumn{2}{c|}{\ding{55}}                    \\ \hline
            Pyro Tube HR    & 3.5$\%$   & 1.8$\%$                  & 2.0$\%$                \\ \hline
        \end{tabular}
        \caption{Average measurement resolution for the procedures, methods and cuvette types used at EGADS and Super-Kamiokande.}
        \label{tab:cuvette-vs-method-errors}
    \end{center}
\end{table}
%%%%%%%%%%%%%%%%%%%%%%%%%%%%%%%%%%%%%%%%%%%%%%%%%%%%%%%%%%%%%%%%%%%%%%%%%%%

Although in principle, the properties of W are better suited for this purpose (melting point, density and expansion coefficient), the Ta platforms yield a lower uncertainty. We think that this could be the result of the machining and molding process, which is more challenging in the case of W.

As mentioned in Section~\ref{sec:intro}, the AAS measurement range is between 5-30 ppm. However, the samples are usually above this range and they must be carefully diluted, their diluted concentration measured and the original concentration calculated from the dilution factor. With these setup and settings, samples from the EGADS and Super-Kamiokande detectors were taken at different depths and measured in~\cite{MARTI} and~\cite{Abe_2022}, respectively. An example of the Gd sulfate concentration for the three sampling positions in the EGADS detector, top, centre and bottom, is shown in Figure 11 in~\cite{MARTI}. Note that in~\cite{MARTI}, we used C HR and HR cuvettes with the La+K doping method because, at that time, we were still developing the Ta/W method. In~\cite{Abe_2022}, we utilized HR cuvettes with Ta/W platforms. The methods used in these references have since undergone further improvements, and uncertainty estimations have been more accurately determined, as described in this paper.

\section{Conclusion}
\label{sec:conclusion}

Two methods for measurements of Gd sulfate with an Atomic Absorption Spectrometer have been presented and compared: doping with La+K and using Ta/W platforms. Additionally, a comparison between the discontinued C HR cuvette and the HR cuvette types when doping with La+K was also shown.

Cuvette type C HR were more stable and showed a smaller variability among cuvettes in the measurements. Moreover, they lasted a larger number of furnace processes which made measurements easier. Despite the discontinued production, we found a good alternative by utilizing Ta and W platforms with the HR cuvettes. Ta and W are two metals that are very suitable for our purposes since they provide mechanical stability and a slow rate of aging  compared to the HR cuvette. Furthermore, they improve atomization of Gd and minimize memory effects. Initially, W appears to be a more promising candidate than Ta. However, we observe a slightly better performance of the Ta platforms compared to the W platforms. This could be due to the difficulty in molding the W pieces, as described in the paper.

The reduced memory effects have another positive effect: less cuvette cleaning is needed and therefore a more efficient usage of the cuvette lifetime can be done. Another positive consequence is the reduction in measurement time to only 25$\%$ of that required by the La+K doping method.

The uncertainty error of a sample measurement for the La+K doping method has been reduced from 3.5$\%$ to 1.8 and 2.0$\%$ with the Ta and W platform methods, respectively. However, the good qualities of the C HR cuvette reminds us the importance of cuvette type availability and choice. It would be interesting to study C HR or of similar quality cuvettes with Ta/W platforms since it could result in an uncertainty reduction and other benefits like signal increase due to increase temperatures, etc.

Both the La+K doping and Ta/W platform methods have been successfully used in assessing the concentration and homogeneity of Gd sulfate in the EGADS and Super-Kamiokande detectors, as well as in monitoring concentration changes.

\section*{Acknowledgments}
We would like to thank the Super-Kamiokande collaboration and very specially to the EGADS group for their help in conducting this study. We also would like to thank the technicians of Kavli-IPMU Nakagawa Hitoshi and Kanazawa Motoichi for their help and support. We gratefully acknowledge the cooperation of the Kamioka Mining and Smelting Company. Ll. Marti has been partially supported by funds from the Ministry of Education, Culture, Sports, Science and Technology (Grant-in-Aid for Young Scientists No. 15K17638). L. Labarga acknowledges the support from the Spanish Ministry of Science, Universities and Innovation (grant PID2021-124050NB-C31) and the European Union's Horizon 2020 Research and Innovation Programme under the Marie Sklodowska-Curie grant agreement no. 872549 H2020-MSCA-RISE-2019 SK2HK.

\vspace{0.5cm}
%\newpage

\bibliographystyle{ptephy}
\bibliography{AAS-paper-body}
%
% once the .bbl file has been generated then place the text in your article.

\end{document}